\documentstyle[12pt]{article}

\begin{document}

\baselineskip=14pt plus 0.2pt minus 0.2pt
\lineskip=14pt plus 0.2pt minus 0.2pt

\begin{flushright}
{}~~~~~~~\\
LA-UR-93-3731 \\
Proceedings of the International Symposium on\\
Coherent States: Past, Present, and Future\\
(to be published)
\end{flushright}

\begin{center}
\Large{\bf
GENERALIZED SQUEEZED STATES FROM GENERALIZED COHERENT STATES} \\

\vspace{0.25in}

\large
\bigskip

Michael Martin Nieto\footnote{Email:  mmn@pion.lanl.gov}\\
{\it
Theoretical Division, Los Alamos National Laboratory\\
University of California\\
Los Alamos, New Mexico 87545, U.S.A. }

\normalsize

\vspace{0.3in}

{ABSTRACT}

\end{center}
\begin{quotation}
Both the coherent states and also the squeezed states of the harmonic
oscillator
have long been understood from the three classical  points of view: the 1)
displacement
 operator, 2) annihilation- (or ladder-) operator, and minimum-uncertainty
methods.
For general systems, there is the same understanding except for
ladder-operator
and displacement-operator  squeezed states.  After reviewing the known
concepts, I propose a  method for obtaining generalized minimum-uncertainty
squeezed states, give examples, and relate it to known concepts.  I comment on
the
remaining  concept, that of general displacement-operator squeezed states.

\end{quotation}

\vspace{0.3in}


\section{Introduction}

As we all know,
 Glauber, Klauder, and Sudarshan  produced
the modern era's seminal works on coherent states of the harmonic oscillator
\cite{5.9,6.1,6,6.2}.
There came to be three classical definitions of these  coherent states, from
1)
{\it the displacement-operator method}, 2) {\it the annihilation-operator
method}
which I will relabel {\it the ladder-operator method}, and 3) {\it the
minimum-uncertainty method}.  Today coherent states are important in many
fields of
theoretical and experimental  physics \cite{1,2}.

Generalizations of these states appeared in two areas.  One was to
``two-photon states" \cite{yuen}, states which were rediscovered under many
names.
In 1979 they were first called  ``squeezed states" \cite{hollen}.  In recent
times
these states  have become of more and more interest \cite{2.5,3}.  This is
especially true in the fields of quantum optics \cite{4} and gravitational wave
detection \cite{5}.  The squeezed states of the harmonic oscillator can also be
equivalently defined by appropriate generalizations of the three classical
definitions of the coherent states.

The other generalization was to non-harmonic oscillator systems.  From the
group
theory or operator point of view, generalized coherent states for Lie groups
were
widely studied from the displacement operator and annihilation operator methods
\cite{6,9.1}.

In 1978 I, along with Mike Simmons and Vincent Gutschick, began a program to
find
the coherent states for arbitrary potentials from a minimum-uncertainty point
of
view.  You give us the potential, be it  Morse, P\"{o}schl-Teller, Coulomb, or
whatever, and
we'll give you the coherent states.  What came as a byproduct was the
realization
that along the way we had also found the
appropriate generalization to the squeezed states.  Further, we found that
 for arbitrary systems the three methods no longer
necessarily gave the same coherent and/or squeezed states.

This all is in the {\bf past}, from the theme of this Symposium. It leaves
appropriate generalizations of squeezed states from the ladder-operator and
displacement-operator points of view to be found.

The {\bf present} deals with work which Rod Truax and I have recently reported
on
\cite{NT}.
(Indeed, in this Proceedings I draw upon much of the contents of Ref.
\cite{NT},
especially in Section 3.)  We proposed a generalization of squeezed states
using the
ladder-operator method. This method was also connected to the
minimum-uncertainty
method and some aspects of special-case displacement-operator squeezed states
which
have been obtained.  It was also found that, as expected, the generalized
squeezed
states from the different methods could be equivalent, but need not be.  (I
note
that connections of these ideas to Rydberg wave packets and other quantum
systems
have been made elsewhere \cite{rydme,rydthem}.)

This leaves the {\bf future}.  What remains to be done is to find generalized
squeezed states from the displacement-operator method.

\section{Past}
\subsection{Coherent states for the harmonic oscillator}

I begin by reviewing the coherent states for the harmonic oscillator.  As is
well-known, there are three standard definitions of these states, which are
equivalent.\\

1) {\it Displacement-Operator Method}. The coherent
 states can be obtained by applying the  unitary displacement operator  on the
ground state \cite{6,6.1}:
\begin{equation}
 |\alpha\rangle \equiv
	D(\alpha)|0\rangle  = \exp[\alpha a^{\dagger} - \alpha^* a] |0\rangle
=\exp\left[-\frac{1}{2}|\alpha|^2\right] \sum_{n} \frac{\alpha ^n}{\sqrt{n!}}
|n\rangle
  ,  \label{D}
\end{equation}
where $|n\rangle$ are the number states.\\

2) {\it Ladder- (or  Annihilation-) Operator Method}.  The
coherent states can also be defined as  the eigenstates of the destruction
operator:
\begin{equation}
a|\alpha\rangle = \alpha |\alpha\rangle.
\end{equation}
This follows from Eq. (\ref{D}), since
\begin{equation}
0 = D(\alpha)a|0\rangle = (a - \alpha)D(\alpha)|0\rangle
=(a-\alpha)|\alpha\rangle.
\end{equation}
These states are the same as the displacement-operator coherent states. \\

3) {\it Minimum-Uncertainty Method}.  This method  harks back to
Schr\"{o}dinger's discovery of the coherent states \cite{sch}.  Recall that
Schr\"{o}dinger wanted to find  states which maintained their shapes and
followed
the classical motion.  For the harmonic oscillator, these are the states which
minimize the uncertainty relation
\begin{equation}
[x,p] = i\hbar, \hspace{0.5in}
(\Delta x)^2(\Delta p)^2 \geq {\frac{1}{4}}{\hbar}^2,
\end{equation}
subject to the constraint that the ground state is a member of the set.  In
wave
function language $(\hbar = \omega =m = 1)$ they are described by
\begin{equation}
\psi_{cs}(x) = [\pi ]^{-1/4}
\exp\left[-\frac{(x-x_0)^2}{2}+ip_0x\right].   \label{cs}
\end{equation}
That the states of Eqs. (\ref{D}) and (\ref{cs}) are the same can be
demonstrated
 by using the generating function for the Hermite polynomials along with the
identifications
\begin{equation}
\sqrt{2}\Re{(\alpha)} = x_0, ~~~~
\sqrt{2}\Im{(\alpha)} = p_0.
\end{equation}
   Observe that of the four original parameters,
$\langle x\rangle$, $\langle x^2 \rangle$, $\langle p\rangle$, and
$\langle p^2\rangle $, only two remain, $\Re{(\alpha)}$ and $\Im{(\alpha)}$.
That
is firstly because the inequality in the uncertainty relation has been
satisfied.
The remaining three parameter set of states is restricted to two parameters
by demanding that the ground state (which corresponds to zero motion) must be a
member of the set.

\subsection{Squeezed states of the harmonic oscillator}

The above three methods also yield equivalent squeezed states for the harmonic
oscillator.

1) {\it Displacement-Operator Method}. In this method one applies the
``squeeze" or SU(1,1) displacement operator on the coherent state:
\begin{equation}
D(\alpha)S(z)|0\rangle = |(\alpha,z)\rangle,
{}~~~~
S(z) = exp[zK_+ - z^*K_-],  \label{Sop}
\end{equation}
where  $K_+ $,
$K_- $,
and $K_0 $
form an su(1,1) algebra amongst themselves:
\begin{equation}
K_+ = \frac{1}{2}a^{\dagger}a^{\dagger}, \hspace{0.5in}
K_- = \frac{1}{2}aa,  \hspace{0.5in}
K_0 = \frac{1}{2}(a^{\dagger}a + \frac{1}{2}), \label{K}
\end{equation}
\begin{equation}
[K_{0},K_{\pm}] = {\pm}K_{\pm}~,~~~~[K_{+},K_{-}] = -2K_{0}.  \label{KK}
\end{equation}
The ordering of $DS$ vs. $SD$ in Eq. (\ref{Sop}) is unitarily equivalent,
amounting
to a change of parameters:
\begin{equation}
D(\alpha)S(z) = S(z)D(\gamma), ~~~
\gamma = \alpha \cosh r~-~\alpha^*e^{i\theta}\sinh r,  \label{sd}
\end{equation}
where $z=re^{i\theta}$.\\

2) {\it Ladder-Operator Method}.  For the harmonic oscillator, this method
again
follows from the displacement-operator method.  Combining the Bogoliubov
transformation \cite{bog},
\begin{equation}
S^{-1}aS = (\cosh r)a ~+~e^{i\theta}(\sinh r)a^{\dagger} ,
\end{equation}
with Eq. (\ref{sd}), one has that
\begin{equation}
\left[ (\cosh r)a ~-~e^{i\theta}(\sinh r)a^{\dagger}\right] |(\alpha,z)\rangle
= \gamma |(\alpha,z)\rangle.     \label{aa}
\end{equation}
One now sees why I relabel this method the ladder-operator method.  For the
squeezed states one needs both the raising (creation) and lowering
(annihilation) operators.\\

3) {\it Minimum-Uncertainty Method}.  From this point of view, the transition
from
coherent to squeezed states is intuitively simple.  These states minimize the
$x-p$ uncertainty relation, without the added restriction that the ground state
(Gaussian) is a member of the set.  That is, these are a {\it three} parameter
set
of states, which are the Gaussians of all widths:
\begin{equation}
\psi_{ss}(x) = [\pi s^2]^{-1/4}
\exp\left[-\frac{(x-x_0)^2}{2s^2}+ip_0x\right],    \label{gg}
\end{equation}
These squeezed states are equivalent to those obtained from the other
formulations.  This can be verified by combining Eqs. (\ref{aa}) and (\ref{gg})
with the relationships
\begin{equation}
x =\frac{(a + a^{\dagger})}{\sqrt{2}}, ~ ~~~
p=\frac{(a - a^{\dagger})}{i\sqrt{2}}.
\end{equation}
The remaining relationships among the  parameters are
\begin{equation}
z = re^{i\theta},  ~~~~ r = \ln{s}.
\end{equation}
(The phase, $\theta$, is an initial time-displacement.)

\subsection{Generalized (displacement- and ladder-operator)\newline  coherent
states}

When one considers  coherent states for general systems one finds that the
coherent
states from all three methods are not, in general, equivalent, although
they may be in particular cases. \\

1) {\it Displacement-Operator Method}. The generalization of this method to
arbitrary
Lie groups has a long history
\cite{6,1,2,9.1}.  (Supersymmetric extensions of it also  exist \cite{10}.)
One simply applies the displacement operator, which is the
unitary
exponentiation of the factor algebra, on to an extremal state.  That is, let
$T$
be a unitary irreducible representation of the group G on a Hilbert space and
let
$|\psi_0\rangle$ be a fixed vector in the space.  Let $G_0$ be the stability
group; i.e.,
\begin{equation}
T(G_0)|\psi_0\rangle = e^{i\alpha(G_0)}|\psi_0\rangle.
\end{equation}
Then
\begin{equation}
|\psi_g\rangle=T(G/G_0)|\psi_0\rangle
\end{equation}
are the coherent states.\\

2) {\it Ladder-Operator Method}.  The
generalization to arbitrary Lie groups is straight forward, and has also been
widely studied \cite{1,2}.  One obtains the eigenstates of the
generalized lowering-operator (assuming there is an extremal state below):
\begin{equation}
A_-|\delta\rangle
 = \delta |\delta\rangle.
\end{equation}

\subsection{Generalized (minimum-uncertainty) coherent and \newline squeezed
states}

In the program I described in the introduction, we wanted to find the coherent
states for arbitrary potential  systems.
We were motivated by the physics of Schr\'{o}dinger.  We wanted
to find states which follow the classical motion as well as possible and
which maintain their shapes as well as possible.  We felt that
meant there was an associated uncertainty relation, but in which variables?
Our
answer was found in a manner completely backwards from the way I now present
it, but
such is often the case with physics.   In fact, along the way, this method also
yields the  squeezed states for general  potential systems
\cite{n1,n2}.

One starts with the classical Hamiltonian  problem and transforms it into the
``natural classical variables," $X_c$ and $P_c$, which vary as the $\sin$ and
the
$\cos$ of the classical $\omega t$.  The  Hamiltonian is therefore of the form
$P_c^2 + X_c^2 $.  One then takes these natural classical variables and
transforms
them into ``natural quantum operators."  Since these are quantum operators,
they
have a commutation relation and uncertainty relation:  \begin{equation}  [X,P]
=
iG, \hspace{0.5in}  (\Delta X)^2(\Delta P)^2 \geq {\frac{1}{4}}\langle G\rangle
^2.
\label{uncert}
\end{equation}
The states
that minimize this uncertainty relation are given by the solutions to the
equation
\begin{equation}
Y\psi_{ss} \equiv
\left(X + \frac{i\langle G\rangle }{2(\Delta P)^2} P\right)\psi_{ss}
=\left(\langle X\rangle +\frac{i\langle G\rangle }{2(\Delta P)^2}\langle
P\rangle
\right)\psi_{ss}.
\end{equation}
Note that of the  four parameters $\langle X\rangle , \langle P\rangle ,
\langle P^2\rangle $, and $\langle G\rangle $, only three are
independent because they satisfy the equality in the uncertainty relation.
Therefore,
\begin{equation}
\left(X + iB P\right)\psi_{ss} = C \psi_{ss}  ,~~~
B = \frac{\Delta X}{\Delta P},  ~~~
 C = \langle X\rangle + i B \langle P\rangle .
\end{equation}
Here $B$ is real and $C$ is complex.  We called these states, $\psi_{ss}(B,C)$,
the  minimum-uncertainty states for general potentials
\cite{n1,n2}.  However, using the  parlance accepted later, they  are  the
squeezed
states for general  potentials \cite{3}.  Then $B$ can be adjusted to $B_0$ so
that
the ground eigenstate of the potential is a member of the set.  Then these
restricted states, $\psi_{ss}(B=B_0,C)=\psi_{cs}(B_0,C)$, are the
minimum-uncertainty coherent states  for general  potentials.

It can be intuitively understood that $\psi_{ss}(B,C)$ and $\psi_{ss}(B_0,C)$
are
the
squeezed and coherent states by recalling the situation for the harmonic
oscillator.
The coherent states are the displaced ground state.  The squeezed states are
Gaussians that have  widths different than that of the ground state
Gaussian, which are
then displaced.

\section{Present: Generalized Ladder-Operator \newline Squeezed States}

As we have discussed, general annihilation-operator (or ladder-operator)
coherent
states are the  eigenstates
of the lowering operator (given a lowest extremal state).
We now propose a generalization to squeezed states, including those for
arbitrary
symmetry
systems:    {\it The general ladder-operator
squeezed states are the eigenstates of a linear combination of the lowering and
raising operators.}

I note that the success of this  method will not be
totally surprising. In many exactly solvable potential systems, the natural
quantum
operators of the minimum-uncertainty method were found to be Hermitian
combinations of the n-dependent raising and lowering operators \cite{n1,n2}.
Here, however, one must generalize to full operators:  $n \rightarrow n(H)$.
Furthermore, in other harmonic-oscillator-like systems, with a Bogoliubov
transformation, this method applies.  (See below.)

I will proceed by
showing how the minimum-uncertainty method for obtaining generalized  squeezed
states can be used as an intuitive tool to aid in understanding the
ladder-operator
method for obtaining generalized squeezed states. I will do this  with two
specific
examples.  Once that is done, the  ladder operator method can be applied to
general
symmetry systems, independent of whether they  come from a
Hamiltonian system in the manner of the minimum-uncertainty method above.
Such is our third example.

\subsection{Example I:  The harmonic oscillator.}

 First let us  re-examine the harmonic
oscillator, starting  from the  minimum-uncertainty method.  Here  $X$ and $P$
are
obviously $x$ and $p$.    Then we have
\begin{equation}
Y = x + s^2 \frac{d}{dx},
\end{equation}
where we have presciently labeled $B$ as $s^2$.  (For the limit to coherent
states,
it turns out that $B=1$.)

Now writing $x$ and $p$ in terms of creation and annihilation operators,
 we find
\begin{equation}
\sqrt{2} \left[
a\left( \frac{1+s^2}{2} \right)
 +  a^{\dagger}
\left( \frac{1-s^2}{2}
\right)
 \right]
\psi_{ss}(s^2,x_0+is^2p_0)
=[x_0+is^2p_0]\psi_{ss}(s^2,x_0+is^2p_0) .
\end{equation}
Therefore, the squeezed states are eigenstates of a linear combination of the
annihilation and creation operators.   Specifically, these states are
 as given in Eq. (\ref{gg}),
\begin{equation}
\psi_{ss}(x) = [\pi s^2]^{-1/4}
\exp\left[-\frac{(x-x_0)^2}{2s^2}+ip_0x\right],   \label{Gauss}
\end{equation}

\subsection{ Example II:  The harmonic oscillator with centripetal barrier.}
I now discuss  the symmetry of the harmonic  oscillator with
centripetal barrier. Previously, the coherent states for this  particular
example
were found with the minimum-uncertainty method, but not the squeezed states
\cite{n2}.  Therefore,  it
is a useful  system since, at the end, we can relate our results to the
coherent
states  obtained
from the minimum-uncertainty method.

This system contains an su(1,1) algebra \cite{truax}. Its elements are
\begin{equation}
L_{\pm} =\frac{1}{4\nu}\frac{d^2}{dz^2}
\mp \frac{1}{2} z \frac{d}{dz} \mp \frac{1}{4} + \frac{\nu}{4} z^2
- \frac{\nu}{4 z^2}  ,
\end{equation}
\begin{equation}
L_0 = \frac{H}{4\nu} + \frac{\nu}{2}  , ~~~~
H = -  \frac{d^2}{dz^2}
+ \nu^2 \left(\frac{1}{z} - z\right)^2 .  \label{L0}
\end{equation}
In terms of the $X$ and $P$ minimum-uncertainty operators \cite{n2}, we find
\begin{equation}
X = \frac{L_- + L_+}{\nu} = z^2 - \left(1+\frac{H}{2\nu^2}\right), ~~~
P = \frac{2(L_- -L_+)}{i} = \frac{1}{i}\left[2z\frac{d}{dz} + 1\right].
\end{equation}
Therefore, the squeezed states for this system are formed by the
solution to the equation
\begin{equation}
0=\left[y\frac{d^2}{dy^2} + \left(\frac{1}{2} + 2\nu By\right)\frac{d}{dy}
+\frac{1}{4}\left(y-\frac{\nu^2}{y}+2B\nu\right)-\frac{\nu C}{2}\right]
\psi_{ss},
\end{equation}
where we have changed variables to $y=\nu z^2$.
The squeezed state solutions to this
equation are
\begin{equation}
\psi_{ss} = N\exp[-y(\nu B + \gamma)]
\left[y^{\lambda + \frac{1}{2}}\right]
\Phi\left(\left[\frac{\nu C}{4\gamma}+\frac{1}{2}(\lambda+\frac{3}{2})
\right],~
\left[\lambda+\frac{3}{2}\right];~2\gamma y\right),
\end{equation}
where $\Phi(a,b;c)$ is the confluent hypergeometric function
$ \sum_{n=0}^{\infty} \frac{(a)_n c^n}{(b)_n~n!}$,
$\gamma =\sqrt{\nu^2 B^2 - \frac{1}{4}} $, and
$ \lambda(\lambda +1) = \nu^2$.
In the limit where $B
\rightarrow 1/(2\nu)$, these become the coherent states given in  Ref.
\cite{n2},
\begin{equation}
\psi_{cs} = \left[\frac{2\nu^{1/2}e^{-\nu \Re(C)}}{I_{\lambda + 1/2}(\nu |C|)}
\right]^{1/2}
e^{-y/2} y^{1/4}I_{\lambda +1/2}\left((2\nu Cy)^{1/2}\right),
\end{equation}
where $I$ is the modified Bessel function. \\

\subsection{ Example III: The squeeze algebra.}  We now consider a symmetry
system
which does not have as its  origin a  Hamiltonian system.  We consider the
su(1,1)
symmetry of Eqs. (\ref{K}, \ref{KK}).  Our ladder-operator squeezed states are
thus
the solutions to  \begin{equation}
\left[\left(\frac{1+s}{2}\right) aa
+\left(\frac{1-s}{2}\right) a^{\dagger} a^{\dagger} \right]\psi_{ss}
=\beta^2 \psi_{ss} .   \label{Kss}
\end{equation}
where the analogue of $B$ is $s$ and the role of $C$ is taken by $\beta^2$.
Using the differential representations of the ladder operators,  Eq.
(\ref{Kss})
can be written in the form
\begin{equation}
 \left[\frac{d^2}{dy^2} +2ys\frac{d}{dy} +y^2 +(s-2\beta^2)\right]
\psi_{ss} = 0.
\end{equation}

Observe that  the ladder operators raise and lower the number states by two
units.
Therefore, there will be two solutions to this equation, one containing only
even
number states and one containing only odd number states.  We will designate
these as
$\psi_{Ess}$ and $\psi_{Oss}$.  These solutions are
\begin{equation}
\psi_{Ess}=N_E\exp{\left[-\frac{-y^2}{2}(s+\sqrt{s^2-1})\right]}
\Phi\left(\left[\frac{1}{4}+\frac{\beta^2}{2\sqrt{s^2-1}}\right],~
\frac{1}{2};~y^2\sqrt{s^2-1}\right),
\end{equation}
\begin{equation}
\psi_{Oss}=N_O ~y\exp{\left[-\frac{-y^2}{2}(s+\sqrt{s^2-1})\right]}
\Phi\left(\left[\frac{3}{4}+\frac{\beta^2}{2\sqrt{s^2-1}}\right],~
\frac{3}{2};~y^2\sqrt{s^2-1}\right).
\end{equation}
In the limit $s\rightarrow 1$, these become the even and odd coherent states:
\begin{equation}
\psi_{Ecs}=
\left[\frac{e^{-\beta^2}}{\pi^{1/2}\cosh{|\beta|^2}}\right]^{1/2}
\exp\left[-\frac{1}{2}y^2\right]\cosh(\sqrt{2}\beta y),
\end{equation}
\begin{equation}
\psi_{Ocs}=
\left[\frac{e^{-\beta^2}}{\pi^{1/2}\sinh{|\beta|^2}}\right]^{1/2}
\exp\left[-\frac{1}{2}y^2\right]\sinh(\sqrt{2}\beta y).
\end{equation}
Using generating formulae, these can be written in the number-state basis as
\begin{equation}
\psi_{Ecs}\ = [\cosh{|\beta|^2}]^{-1/2}
\sum_{n=0}^{\infty}\frac{\beta^{2n}}{\sqrt{(2n)!}}|2n\rangle ,
\end{equation}
\begin{equation}
\psi_{Ocs}\ = [\sinh{|\beta|^2}]^{-1/2}
\sum_{n=0}^{\infty}\frac{\beta^{2n+1}}{\sqrt{(2n+1)!}}|2n+1\rangle .
\end{equation}
Up to the normalization, these are the ``even and odd coherent states"
previously
found in Ref. \cite{mankoEO}.
Although this system did not come from a Hamiltonian, one could have
used a minimum-uncertainty principle to obtain the same states by starting with
the commutation relation
$[K_{+},K_{-}] = -2K_{0}$.
However, one does not obtain the same coherent states from
the displacement-operator method.  Those coherent states, defined by
$S(z)|0\rangle$,   are  the squeezed-vacuum Gaussian of Eq. (\ref{Gauss}) with
$x_0 = p_0 = 0$.  \\

The above three examples have all been cases where
$A_- = (A_+)^{^{\dagger}}$.  Sometimes that is not the case, as in certain
potential systems
whose eigenenergies are not equally spaced \cite{n1,n2}.  Then, as in Eq.
(\ref{L0}), one should use the operator form for ``$n$": $A_n \rightarrow
A_{n(H)}$, to connect to the minimum-uncertainty method.  In these cases, the
ladder-operator coherent and squeezed states can be different than, though
related
to, their minimum-uncertainty counterparts.

\section{Future:  Generalized Displacement-Operator \newline Squeezed States?}

Although the displacement-operator method is the natural one for defining
coherent states of Lie groups,  there is
as yet no well-known general extension of this method to define general
displacement-operator squeezed
states.  It  has basically only been
applied to  harmonic-oscillator-like systems
 \cite{2.5,3}.

This  has been touched upon in discussions \cite{sq1} about
higher-order generalizations of the ``squeeze operator," $S(z)$.  In
particular,
although harmonic-oscillator-like systems admit squeeze operators (or
Bogoliubov
transformations) connecting  the  displacement-operator and ladder-operator
methods
\cite{ass,Agar}, the appropriate generalization of these squeeze operators have
not
been found.  Therefore, for now, the ladder-operator method is generally
connected
only to the minimum-uncertainty method.

Note also that for finite-dimensional representations, such
as for angular momentum coherent states, the ladder-operator method does not
allow a
solution for coherent states, although the displacement-operator method does
\cite{ass}.  Contrariwise, for squeezed states, we observe that the opposite is
true.

It thus remains for the future to find generalized squeezed states from the
displacement operator method.  I have been given some helpful advice by a
number
of people at this Symposium, including John Klauder and Arthur Wightman, but
the
solution remains to be found.  Hopefully, at the next Symposium something
further
can be said.

\end{document}